\documentclass[3p,times,twocolumn]{elsarticle}

\usepackage{ecrc,amsmath}

\volume{00}

\firstpage{1}

\journalname{Nuclear Physics B Proceedings Supplement}

\runauth{}

\jid{nuphbp}

\jnltitlelogo{Nuclear Physics B Proceedings Supplement}

\usepackage{amssymb}

\usepackage[figuresright]{rotating}

\newcommand{ \be}{\begin{equation}}
\newcommand{ \ee}{\end{equation}}
\newcommand{ \bea}{\begin{eqnarray}}
\newcommand{ \eea}{\end{eqnarray}}
\newcommand{ \mysmall}[1]{\scriptscriptstyle #1} 

\def\eq#1{Eq.~(\ref{#1})}

\begin{document}

\begin{frontmatter}



\dochead{}

\title{Towards a determination of the tau lepton dipole moments}


\author[a,b]{M.\ Fael}
\author[c]{L.\ Mercolli}
\author[d]{M.\ Passera}

\address[a]{Dipartimento di Fisica e Astronomia, Universit\`a di Padova, 35131 Padova, Italy}
\address[b]{Institut f\"{u}r Theoretische Physik, Universit\"{a}t Z\"{u}rich, CH-8057, Z\"{u}rich, Switzerland}
\address[c]{Princeton University, Department of Astrophysical Sciences, Princeton, NJ, 08544, USA}
\address[d]{Istituto Nazionale Fisica Nucleare, 35131 Padova, Italy}

\begin{abstract} 
The $\tau$ anomalous magnetic moment $a_{\tau}$ and electric dipole moment $d_{\tau}$ have not yet been observed. The present bounds on their values are of order $10^{-2}$ and $10^{-17} e$ cm, respectively. We propose to measure $a_{\tau}$ with a precision of $O(10^{-3})$ or better and improve the existing limits on $d_{\tau}$ using precise $\tau^- \to l^- \nu_\tau \, \bar{\nu}_l\, \gamma$ $(l=e$ or $\mu$) data from high-luminosity $B$ factories. A detailed feasibility study of this method is underway.
\end{abstract}



\end{frontmatter}



\section{Introduction}

The very short lifetime of the $\tau$ lepton ($2.9 \times 10^{-13}$s) makes it very difficult to measure its electric and magnetic dipole moments. The present resolution on its anomalous magnetic moment $a_{\tau}$ is only of $O(10^{-2})$~\cite{Abdallah:2003xd}, more than an order of magnitude larger than its Standard Model (SM) prediction $\frac{\alpha}{2\pi} \simeq 0.001$. In fact, while the SM value of $a_{\tau}$ is known with a tiny uncertainty of $5 \times 10^{-8}$~\cite{Eidelman:2007sb}, this short lifetime has so far prevented the determination of $a_{\tau}$ measuring the $\tau$ spin precession in a magnetic field, like in the electron and muon $g$$-$$2$ experiments. Instead, experiments focused on various high-precision measurements of $\tau$ pair production in high-energy processes, comparing the measured cross sections with the SM predictions. As these processes involve off-shell photons or taus in the $\tau \bar{\tau} \gamma$ vertices, the measured quantity is not directly $a_{\tau}$.

A precise measurement of $a_{\tau}$ would offer an excellent opportunity to unveil new physics effects. Indeed, in a large class of theories beyond the SM, new contributions to the anomalous magnetic moment of a lepton $l$ of mass $m_l$ scale with $m_l^2$. Therefore, given the large factor $m_{\tau}^2/m_{\mu}^2 \sim 283$, the $g$$-$$2$ of the $\tau$ is much more sensitive than the muon one to electroweak and new physics loop effects that give contributions proportional to $m_l^2$. In these scenarios, the present discrepancy in the muon $g$$-$$2$ suggests a new-physics effect in $a_{\tau}$ of $O(10^{-6})$; several theories exist where this na\"{\i}ve scaling is violated and much larger effects are expected~\cite{Giudice:2012ms}.

Ever since the very first discovery of $CP$ violation in the 1960s there have been ongoing efforts to measure fundamental electric dipole moments (EDMs). Indeed, EDMs violate parity and time reversal; if $CPT$ is a good symmetry, $T$ violation implies $CP$ violation and vice versa. 
The SM predictions for lepton EDMs are far too small to be seen by projected future experiments. Hence, the observation of a non-vanishing lepton EDM would be evidence for a $CP$-violating new physics effect~\cite{EDM_General}.
The experimental determination or a stringent bound on the $\tau$ lepton EDM $d_{\tau}$ poses the same difficulties that one encounters in the case of $a_\tau$: the $\tau$'s lifetime is very short. Nonetheless, a $CP$ violation signature arising from $d_{\tau}$ was searched for in the $e^+ e^- \to \tau^+ \tau^-$ reaction reaching a sensitivity to $d_{\tau}$ of $10^{-17} e$ cm~\cite{Inami2002}.

After a brief review of the present experimental and theoretical status of the $\tau$ dipole moments, we discuss the possibility to probe them via precise measurements of the decays 
$\tau^- \to l^- \nu_\tau \, \bar{\nu}_l\, \gamma$
($l=e$ or $\mu$).

\section{The anomalous magnetic moment of the tau}


The SM prediction for $a_{\tau}$ is given by the sum of QED, electroweak (EW) and hadronic terms. The QED contribution has been computed up to three loops:
$
    a_{\tau}^{\mysmall \rm QED} =
    117 \, 324 \, (2) \times 10^{-8}
$
(see~\cite{Passera:2006gc} and references therein), where the uncertainty
$\pi^2 \ln^2(m_{\tau}/m_e)(\alpha/\pi)^4 \sim 2\times 10^{-8}$
has been assigned for uncalculated four-loop contributions. The errors due to the uncertainties of the $O(\alpha^2)$ ($5 \times 10^{-10}$) and $O(\alpha^3)$ terms ($3 \times 10^{-11}$), as well as that induced by the uncertainty of $\alpha$ ($8 \times 10^{-13}$), are negligible.
The sum of the one- and two-loop EW contributions is
$
    a_{\tau}^{\mysmall \rm EW} = 47.4 (5) \times 10^{-8}
$
(see \cite{ew2loop,Eidelman:2007sb} and references therein). The uncertainty encompasses the estimated errors induced by hadronic loop effects, neglected two-loop bosonic terms and the missing three-loop contribution. It also includes the tiny errors due to the uncertainties in $M_{\rm\scriptstyle top}$ and $m_{\tau}$.

Similarly to the case of the muon $g$$-$$2$, the leading-order hadronic contribution to $a_{\tau}$ is obtained via a dispersion integral of the total hadronic cross section of the $e^+e^-$ annihilation (the role of low energies is very important, although not as much as for $a_{\mu}$). The result of the latest evaluation, using the whole bulk of experimental data below 12~GeV, is
$
    a_{\tau}^{\mysmall \rm HLO} =  337.5 \, (3.7) \times 10^{-8}
$~\cite{Eidelman:2007sb}.
The hadronic higher-order $(\alpha^3)$ contribution $a_{\tau}^{\mysmall \rm HHO}$ can be divided into two parts:
$
     a_{\tau}^{\mysmall \rm HHO}=
     a_{\tau}^{\mysmall \rm HHO}(\mbox{vp})+
     a_{\tau}^{\mysmall \rm HHO}(\mbox{lbl}).
$
The first one, the $O(\alpha^3)$ contribution of diagrams containing hadronic self-energy insertions in the photon propagators,
is
$
a_{\tau}^{\mysmall \rm HHO}(\mbox{vp})= 7.6 (2) \times 10^{-8}
$~\cite{Krause:1996rf}.
Note that na\"{\i}vely rescaling the corresponding muon $g$$-$$2$ result by a factor $m_{\tau}^2/m_{\mu}^2$ leads to the incorrect estimate $a_{\tau}^{\mysmall \rm HHO}(\mbox{vp}) \sim -28\times 10^{-8}$ (even the sign is wrong!). 
Estimates of the light-by-light contribution $a_{\tau}^{\mbox{$\scriptscriptstyle{\rm HHO}$}}(\mbox{lbl})$ obtained rescaling the corresponding one for the muon $g$$-$$2$ by a factor $m_{\tau}^2/m_{\mu}^2$ fall short of what is needed -- this scaling is not justified. The parton-level estimate of~\cite{Eidelman:2007sb} is
$
a_{\tau}^{\mysmall \rm HHO}(\mbox{lbl})= 5 (3) \times 10^{-8},
$
a value much lower than those obtained by na\"{\i}ve rescaling. Adding up the above contributions one obtains the SM 
prediction~\cite{Eidelman:2007sb}
\be
    a_{\tau}^{\mysmall \rm SM} = 
         a_{\tau}^{\mysmall \rm QED} +
         a_{\tau}^{\mysmall \rm EW}  +
         a_{\tau}^{\mysmall \rm HLO}  +
         a_{\tau}^{\mysmall \rm HHO}
         =117 \, 721 \, (5) \times 10^{-8}.  
\label{eq:atSM}
\ee
Errors were added in quadrature.


The present PDG limit on the $\tau$  lepton $g$$-$$2$ was derived in 2004 by the DELPHI collaboration from $e^+e^- \to e^+e^-\tau^+\tau^-$ total cross section measurements at $\sqrt{s}$ between 183 and 208 GeV at LEP2 (the study of $a_{\tau}$ via this channel was proposed in~\cite{Cornet:1995pw}):
$
                         -0.052 < a_{\tau} < 0.013
$
at $95\%$ confidence level~\cite{Abdallah:2003xd}. Reference~\cite{Abdallah:2003xd} also quotes the result in the form:
\be
                         a_{\tau} = -0.018 (17).
\label{eq:atEXP}
\ee
This limit was obtained comparing the experimentally measured values of the cross section to the SM values, assuming that possible deviations were due to non-SM values of $a_{\tau}$. Therefore, Eq.~(\ref{eq:atEXP}) is actually a limit on the non-SM contribution to $a_{\tau}$.
We refer to~\cite{Lohmann:2005im} for a concise review of older, less stringent limits on $a_{\tau}$, derived from data samples of  $e^+e^- \to \tau^+\tau^-$ (for photon squared four-momentum $q^2$ up to $\mbox{(37 GeV)}^2$)~\cite{Silverman:1982ft} and $Z^0 \to \tau^+\tau^- \gamma$ (where $q^2\!=\!0$ but the radiating $\tau$ is not on-shell)~\cite{Grifols:1990ha,Ackerstaff:1998mt,Acciarri:1998iv,Biebel:1996ur}. Other indirect bounds on $a_{\tau}$ were studied in~\cite{EscribanoMasso}.

Comparing Eqs.~(\ref{eq:atSM}) and (\ref{eq:atEXP}) (their difference is roughly one standard deviation), it is clear that the sensitivity of the best existing measurements is still more than one order of magnitude worse than needed. The possibility to improve these limits is certainly not excluded. Future high-luminosity $B$ factories such as Super-KEKB~\cite{Aushev:2010bq} offer new opportunities to improve the determination of the $\tau$ magnetic properties. 
The authors of \cite{Bernabeu:2007-8} proposed to determine the Pauli form factor $F_2(q^2)$ of the $\tau$ via $\tau^+\tau^-$ production in $e^+ e^-$ collisions at the $\Upsilon$ resonances with sensitivities possibly down to $O(10^{-5})$ or even better.
In \cite{GonzalezSprinberg:2000mk} the reanalysis of various measurements of the cross section of the process $e^+e^- \to \tau^+\tau^-$, the transverse $\tau$ polarization and asymmetry at LEP and SLD, as well as the decay width $\Gamma(W \to \tau\nu_{\tau})$ at LEP and Tevatron allowed the authors to set a model-independent limit on new physics contributions,
\be
                       -0.007 < a_{\tau}^{\mysmall \rm NP}  < 0.005,
\ee
a bound stronger than that in Eq.~(\ref{eq:atEXP}). This analysis, like earlier ones, was performed without radiative corrections, but the authors checked that the inclusion of initial-state radiation did not affect significantly the obtained bounds.

\section{The $\tau$ lepton EDM}

In the SM, the only sources of $CP$ violation are the CKM-phase and a possible $\theta$-term in QCD. A fundamental EDM can only be generated at the three-loop level \cite{EDM-3loop}, yielding a SM contribution which is far below experimental capabilities. Models for physics beyond the SM generally induce large contributions to lepton and neutron EDMs, and although there has been no experimental evidence for an EDM so far, there is considerable hope to gain new insights into the nature of $CP$ violation through this kind of experiments.

The current $95\%$ confidence level limits on the EDM of the $\tau$ lepton are given by 
\begin{equation}\label{eq dtauexp}
\begin{split}
& - 2.2 < \mathrm{Re} (d_\tau) < 4.5 \; \; (10^{-17} \;  e \, \mathrm{cm}), \\
& - 2.5 < \mathrm{Im} (d_\tau) < 0.8 \; \; (10^{-17} \;  e \, \mathrm{cm}); \\
\end{split}
\end{equation}
they were obtained by the Belle collaboration \cite{Inami2002} following the analysis of \cite{Nachtmann1993} for the impact of an effective operator for the $\tau$ EDM in the process $e^+ e^- \rightarrow \tau^+ \tau^-$. Compared to $a_\tau$, the experimental sensitivity to $d_\tau$ is not significantly higher since in natural units the bounds on $d_\tau$ quoted above are of the order of $10^{-3} \; \mathrm{GeV}^{-1}$. Furthermore, as discussed previously in the case of $a_\tau$, the analysis of \cite{Inami2002} is limited to the LO precision of the calculation of \cite{Nachtmann1993}. Therefore, we believe that there is considerable room for an improvement of these bounds.

\section{Radiative leptonic $\tau$ decays}

We propose to measure the dipole moments of the $\tau$ lepton through its radiative leptonic decays \cite{Matteo, us, LSS}
\begin{equation}
   \tau^- \to l^- \nu_\tau \, \bar{\nu}_l\, \gamma, \quad \mbox{with } l=e \;  \mbox{or} \; \mu.
   \label{eq:leptonicradiativedecays}
\end{equation}
The authors of \cite{GonzalezSprinberg:2000mk} and \cite{Nachtmann1993} have applied effective Lagrangian techniques to study $a_\tau$ and $d_\tau$. Our strategy is similar: we describe radiative $\tau$ decays through an effective Lagrangian $\mathcal{L}_\text{eff}$ which contains the QED Lagrangian for the leptons, the effective Fermi Lagrangian, and the effective operators that contribute to the anomalous magnetic moment and EDM of the $\tau$ lepton, i.e.\
\begin{equation}
\mathcal{L}_\text{eff} =   \mathcal{L}_\text{QED} + \mathcal{L}_\text{Fermi}  + c_a  \frac{e}{4 \Lambda} \; \mathcal{O}_a   - c_d \frac{i}{2 \Lambda} \; \mathcal{O}_d, 
\label{eq:leff}
\end{equation}
where $\mathcal{O}_{a,d}$ are given by 
\begin{equation}
\begin{split}
& \mathcal{O}_{a} \; = \;  \bar{\tau} \sigma_{\mu \nu}  \tau  \, F^{\mu \nu}, \\
& \mathcal{O}_{d} \; =  \; \bar{\tau} \sigma_{\mu \nu} \gamma_5 \tau  \, F^{\mu \nu}.
\end{split}
\label{eq:ops}
\end{equation}
The scale $\Lambda$ represents the scale where any kind of physics which is not described by $\mathcal{L}_\text{eff}$ generates a contribution to the $\tau$'s electric or magnetic dipole moment and is therefore larger than the electroweak scale, i.e.\ $\Lambda > M_Z$. For simplicity we assume the scale $\Lambda$ to be equal for both operators $\mathcal{O}_{a,d}$, knowing that in fact the scale for the EDM is much higher than that for the $g$$-$$2$. The effective Lagrangian $\mathcal{L}_\text{eff}$ in \eq{eq:leff} gives the following predictions for the $\tau$ dipole moments: 
\bea
a_\tau  & = &    \frac{\alpha}{2\pi} + c_a  \frac{m_\tau }{ \Lambda } + \cdots	
   \label{eq:atauTH}
\\
d_\tau   & =  & c_d \, \frac{1}{\Lambda} + \cdots
   \label{eq:dtauTH}
\eea
where the dots indicate higher-order contributions not relevant for our discussion (note, in \eq{eq:dtauTH}, that $d_{\tau}$ has no QED contribution). We then define the parameters
\bea
\tilde{a}_\tau  & \equiv &   c_a  \frac{m_\tau }{ \Lambda },	\\
\tilde{d}_\tau   & \equiv  & c_d \, \frac{1}{\Lambda},
\eea
which can be determined as follows.

The leading-order (LO) prediction of the Lagrangian $\mathcal{L}_\text{QED} + \mathcal{L}_\text{Fermi}$ for the differential decay rate of a polarized $\tau^-$ into $l^- \!  + \bar{\nu}_l  + \nu_{\tau} + \gamma$ ($l=e$ or $\mu$) is, in the $\tau$ lepton rest frame,
\begin{multline}
	\frac{d\Gamma_0^6}{dx \, dy \, d \Omega_l \, d\Omega_{\gamma}} = 
	\frac{\alpha \, m_{\tau}^5 \, G_F^2}{(4 \pi)^8}  \, y \sqrt{x^2 -4r^2}  \,\, \Bigg[G_0(x,y,c) 
		\\
	+ \! \sqrt{x^2-4r^2}  J_0(x,y,c)  \, \vec{n}\cdot \hat{p}_l 
	+ y  K_0(x,y,c) \, \vec{n} \cdot \hat{p}_\gamma \Bigg],
   \label{eq:decayrate}
\end{multline}
where $\alpha = 1/137.035\,999\,174\,(35)$~\cite{Aoyama:2012wj} is the fine-structure constant, $G_F=1.663788(7) \times 10^{-5}$ GeV$^{-2}$~\cite{Webber:2010zf} is the Fermi coupling constant, $m_\tau = 1.77682\,(16)$ GeV \cite{Beringer:1900zz}, $r=m_l/m_\tau$, and the kinematic variables
\begin{equation}
   x=\frac{2E_l}{m_\tau},\quad
   y=\frac{2E_\gamma}{m_\tau}, \quad
   c \equiv \cos \theta
\end{equation}
are the normalized energies of the lepton $l^-$ and the photon, which are respectively emitted at solid angles $\Omega_l$ and  $\Omega_{\gamma}$, and $\theta$ is the angle between their momenta $\vec{p}_l$ and $\vec{p}_\gamma$. Also, $n = (0,\vec{n})$ is the $\tau$ polarization vector with $n^2=-1$ and $n \cdot p_\tau=0$.
The functions $G_0$, $J_0$ and $K_0$ were computed in~\cite{KS1959,Fronsdal:1959zzb,Kuno:1999jp,Fischer:1994pn}.

The operator $\mathcal{O}_{a}$ generates additional contributions to the differential decay rate in~\eq{eq:decayrate}. They can be summarised in the shift~\cite{LSS}
\be
	G_0(x,y,c)   \,\to\,   G_0(x,y,c)   \,+\,   \tilde{a}_\tau  G_a(x,y,c),
\label{eq:achange}
\ee
and similarly for $J_0$ and $K_0$. The operator $\mathcal{O}_{d}$ produces the additional term
\be
\tilde{d}_\tau \, m_{\tau} \, y \sqrt{x^2 - 4 r^2} \,   \hat{p}_l
\cdot \left(\hat{p}_\gamma \times \vec{n} \right)\,  G_d(x,y,z)
\label{eq:dchange}
\ee
inside the square brackets of~\eq{eq:decayrate}~\cite{us}. Tiny terms of $O(\tilde{a}^2)$ and $O(\tilde{d}^2)$ were neglected.

Our goal is to provide a method to determine $a_\tau$ with a precision of $O(10^{-3})$ or better. This calls for an analogous precision on the theoretical side. For this reason, we have corrected the decay rate in Eq.~\eqref{eq:decayrate} to include radiative corrections at next-to-leading order (NLO) in QED~\cite{Matteo,us,Fischer:1994pn,Arbuzov:2004wr}. The comparison of this NLO prediction, modified by the additional terms in \eq{eq:achange} for $G_0$ (and similarly for $J_0$ and $K_0$), to sufficiently precise data allows to determine $\tilde{a}_\tau$ (and thereby $a_\tau$ via \eq{eq:atauTH}) possibly down to the level of $O(10^{-4})$. Similarly, including the additional term in \eq{eq:dchange}, one can also determine $\tilde{d}_\tau$ which, at this level of precision,  coincides with $d_\tau$ (see \eq{eq:dtauTH}). This analysis is in progress~\cite{us}.

The contributions from the two effective operators $\mathcal{O}_{a,d}$ to the electromagnetic form factors are the same for $q^2 = 0$ as for $q^2 \neq 0$. The point is that only higher dimensional operators would give rise to a difference between these two cases, which means that such contributions are suppressed by higher powers of $q^2 / \Lambda^2$~\cite{GonzalezSprinberg:2000mk}. In our case, $q^2$ may be of the order of $m_\tau^2$ while $\Lambda$ is certainly higher than $M_Z$ and we may therefore safely neglect contributions from higher dimensional operators. Of course, the requirement that $q^2 \ll \Lambda^2$ is the fundamental hypothesis of our effective Lagrangian approach.

The possibility to set bounds on $a_{\tau}$ via the radiative leptonic $\tau$ decays in \eq{eq:leptonicradiativedecays} was suggested long ago in Ref.~\cite{LSS}. In that article the authors proposed to take advantage of a radiation zero of the LO differential decay rate in \eq{eq:decayrate} which occurs when, in the $\tau$ rest frame, the final lepton $l$ and the photon are back-to-back, and $l$ has maximal energy. Since a non-standard contribution to $a_{\tau}$ spoils this radiation zero, precise measurements of this phase-space region could be used to set bounds on its value. However, this method is only sensitive to large values of $a_{\tau}$ (at the radiation zero the dependence on non-standard $a_{\tau}$ contributions is quadratic), and preliminary studies with Belle data show no significant improvement of the existing limits~\cite{us}.

\section{Conclusions}

The magnetic and electric dipole moments of the $\tau$ lepton are largely unknown. We propose to use radiative leptonic $\tau$ decays to measure the former with a sensitivity of $O(10^{-3})$ or better and improve the existing limits on the latter. A preliminary analysis with Belle data is in progress to determine the feasibility of this proposal~\cite{us}.

\section*{Acknowledgements}

We would like to thank Denis Epifanov and Simon Eidelman for our very fruitful collaboration, and Brando Bellazzini and Ferruccio Feruglio for very useful discussions.
The work of M.F.\ is supported in part by the Research Executive Agency of the European Union under the Grant Agreement  PITN-GA-2010-264564 (LHCPhenoNet).
L.M.\ is supported by a grant from the Swiss National Science Foundation. 
M.P.\ also thanks the Department of Physics and Astronomy of the University of Padova for its support. His work was supported in part by the Italian Ministero dell'Universit\`a e della Ricerca Scientifica under the program PRIN 2010-11, and by the European Programmes UNILHC (contract PITN-GA-2009-237920) and INVISIBLES (contract PITN-GA-2011-289442).





\begin{thebibliography}{00}


\bibitem{Abdallah:2003xd}
  J.~Abdallah {\it et al.}  [DELPHI Collaboration],
  Eur.\ Phys.\ J.\ C {\bf 35} (2004) 159.

\bibitem{Eidelman:2007sb}
  S.~Eidelman and M.~Passera,
  Mod.\ Phys.\ Lett.\ A {\bf 22} (2007) 159.

\bibitem{Giudice:2012ms}
  G.F.~Giudice, P.~Paradisi and M.~Passera,
  JHEP {\bf 1211} (2012) 113.
  
\bibitem{EDM_General}
  M.~Raidal,  {\it et al.},
  Eur.\ Phys.\ J.\ C {\bf 57} (2008) 13;
  M.~Pospelov and A.~Ritz,
  in {\it ``Lepton dipole moments,''}
  B.L.~Roberts and W.J.~Marciano (eds.),
  Advanced series on directions in high energy physics, Vol.\ 20, World Scientific (2010), p.\ 439;
  %
  L.~Mercolli and C.~Smith,
  Nucl.\ Phys.\ B {\bf 817} (2009) 1.

\bibitem{Inami2002}
  K.~Inami {\it et al.} [Belle Collaboration],
  Phys.\ Lett.\  {\bf B551 } (2003) 16. 
  
\bibitem{Passera:2006gc}
  M.~Passera,
  Phys.\ Rev.\ D {\bf 75} (2007) 013002.

\bibitem{ew2loop}   A.~Czarnecki, B.~Krause and W.J.~Marciano,
	Phys.\ Rev.\ D {\bf 52} (1995) 2619;
	Phys.\ Rev.\ Lett.\  {\bf 76} (1996) 3267;
	T.V.~Kukhto, E.A.~Kuraev, Z.K.~Silagadze, A.~Schiller,
                  Nucl.\ Phys.\ B {\bf 371} (1992) 567.
       
\bibitem{Krause:1996rf}
  B.~Krause,
  Phys.\ Lett.\ B {\bf 390} (1997) 392.

\bibitem{Cornet:1995pw}
  F.~Cornet and J.I.~Illana,
  Phys.\ Rev.\ D {\bf 53} (1996) 118.
  
  \bibitem{Lohmann:2005im}
  W.~Lohmann,
  Nucl.\ Phys.\ Proc.\ Suppl.\  {\bf 144} (2005) 122.
  
   \bibitem{Silverman:1982ft}
  D.J.~Silverman and G.L.~Shaw,
  Phys.\ Rev.\ D {\bf 27} (1983) 1196.
  
  \bibitem{Grifols:1990ha}
  J.A.~Grifols and A.~Mendez,
  Phys.\ Lett.\ B {\bf 255} (1991) 611
   [Erratum-ibid.\ B {\bf 259} (1991) 512].
    
    \bibitem{Ackerstaff:1998mt}
  K.~Ackerstaff {\it et al.}  [OPAL Collaboration],
  Phys.\ Lett.\ B {\bf 431} (1998) 188.
  
  \bibitem{Acciarri:1998iv}
  M.~Acciarri {\it et al.}  [L3 Collaboration],
  Phys.\ Lett.\ B {\bf 434} (1998) 169.

\bibitem{Biebel:1996ur}
  J.~Biebel and T.~Riemann,
  Z.\ Phys.\ C {\bf 76} (1997) 53.
  
\bibitem{EscribanoMasso}
  R.~Escribano and E.~Masso,
  Phys.\ Lett.\ B {\bf 301} (1993) 419;
  Phys.\ Lett.\ B {\bf 395} (1997) 369.

\bibitem{Aushev:2010bq}
  T.~Aushev {\it et al.},
  arXiv:1002.5012 [hep-ex].
  
\bibitem{Bernabeu:2007-8}
  J.~Bernabeu, G.A.~Gonzalez-Sprinberg, J.~Papavassiliou and J.~Vidal,
  Nucl.\ Phys.\ B {\bf 790} (2008) 160;
 J.~Bernabeu, G.A.~Gonzalez-Sprinberg and J.~Vidal,
  JHEP {\bf 0901} (2009) 062.

\bibitem{GonzalezSprinberg:2000mk}
  G.A.~Gonzalez-Sprinberg, A.~Santamaria and J.~Vidal,
  Nucl.\ Phys.\ B {\bf 582} (2000) 3.

\bibitem{EDM-3loop}
  J.F.~Donoghue,
  Phys.\ Rev.\  {\bf D18 } (1978) 1632;
  E.P.~Shabalin,
  Sov.\ J.\ Nucl.\ Phys.\  {\bf 28 } (1978)  75.
  
\bibitem{Nachtmann1993}
  W.~Bernreuther, O.~Nachtmann, and P.~ Overmann, 
  Phys.\ Rev.\  {\bf D48} (1993) 78.

\bibitem{Matteo}
 M.~Fael, {\it Study of the anomalous magnetic moment of the $\tau$ lepton via its radiative leptonic decays}, 
 M.Sc.\ Thesis (2010),  University of Padua.  
 
 \bibitem{us}
 S.~Eidelman, D.~Epifanov, M.~Fael, L.~Mercolli, and M.~Passera, in preparation.  
 
\bibitem{LSS}
M.L.~Laursen, M.A.~Samuel and A.~Sen,
 Phys.\ Rev.\ D {\bf 29} (1984) 2652 [Erratum-ibid.\ D {\bf 56} (1997) 3155].

\bibitem{Aoyama:2012wj}
  T.~Aoyama, M.~Hayakawa, T.~Kinoshita and M.~Nio,
  Phys.\ Rev.\ Lett.\  {\bf 109} (2012) 111807.
    
\bibitem{Webber:2010zf}
  D.M.~Webber {\it et al.}  [MuLan Collaboration],
  Phys.\ Rev.\ Lett.\  {\bf 106} (2011) 041803.
    
\bibitem{Beringer:1900zz}
  J.~Beringer {\it et al.}  [Particle Data Group Collaboration],
  Phys.\ Rev.\ D {\bf 86} (2012) 010001.

\bibitem{KS1959}
  T.~Kinoshita and A.~Sirlin,
  Phys.\ Rev.\ Lett.\  {\bf 2} (1959) 177.
  
\bibitem{Fronsdal:1959zzb}
  C.~Fronsdal and H.~Uberall,
  Phys.\ Rev.\  {\bf 113} (1959) 654.

\bibitem{Kuno:1999jp}
  Y.~Kuno and Y.~Okada,
  Rev.\ Mod.\ Phys.\  {\bf 73} (2001) 151.

\bibitem{Fischer:1994pn}
  A.~Fischer, T.~Kurosu and F.~Savatier,
  Phys.\ Rev.\ D {\bf 49} (1994) 3426.
  
\bibitem{Arbuzov:2004wr}
  A.B.~Arbuzov and E.S.~Scherbakova,
  Phys.\ Lett.\ B {\bf 597} (2004) 285.




\end{thebibliography}



\end{document}